\hsize=31pc 
\vsize=49pc 
\lineskip=0pt 
\parskip=0pt plus 1pt 
\hfuzz=1pt   
\vfuzz=2pt 
\pretolerance=2500 
\tolerance=5000 
\vbadness=5000 
\hbadness=5000 
\widowpenalty=500 
\clubpenalty=200 
\brokenpenalty=500 
\predisplaypenalty=200 
\voffset=-1pc 
\nopagenumbers      
\catcode`@=11 
\newif\ifams 
\amsfalse %\amstrue si ¤ suivant utilise
% 
%%%%%%%%%%%%%%%%%%%%%%%%%%%%%%%%%%%%%%%%%%%%%%%%%%%%%%%%%%%%% 
%                                                           % 
%  The following section may be commented out and           % 
%  \ifams set to either \amstrue to use the AMS fonts       % 
%  or \amsfalse if they are not available                   % 
%                                                           % 
%%%%%%%%%%%%%%%%%%%%%%%%%%%%%%%%%%%%%%%%%%%%%%%%%%%%%%%%%%%%% 
% 
%\def\Yesreply{Y } 
%\def\Noreply{N } 
%\def\yesreply{y } 
%\def\noreply{n } 
%\newif\ifnotyorn 
%\message{Do you want to use AMSfonts, msam and msbm? Y or N: }% 
%\loop 
%\read-1 to \reply 
%\ifx\reply\yesreply\global\amstrue\notyornfalse 
%\else\ifx\reply\Yesreply\global\amstrue\notyornfalse 
%\else\ifx\reply\noreply\global\amsfalse\notyornfalse 
%\else\ifx\reply\Noreply\global\amsfalse\notyornfalse 
%\else\notyorntrue 
%\message{Please type y or Y  (Yes) or n or N (No)}\fi\fi\fi\fi 
%\ifnotyorn\repeat 
%%%%%%%%%%%%%%%%%%%%%%%%%%%%%%%%%%%%%%%%%%%%%%%%%%%%%%%%%%%% 
% 
\newfam\bdifam 
\newfam\bsyfam 
\newfam\bssfam 
\newfam\msafam 
\newfam\msbfam 
\newif\ifxxpt    
\newif\ifxviipt  
\newif\ifxivpt   
\newif\ifxiipt   
\newif\ifxipt    
\newif\ifxpt     
\newif\ifixpt    
\newif\ifviiipt  
\newif\ifviipt   
\newif\ifvipt    
\newif\ifvpt     
% 
% Headings in 20pt, 17pt or 14pt 
% 
\def\headsize#1#2{\def\headb@seline{#2}% 
                \ifnum#1=20\def\HEAD{twenty}% 
                           \def\smHEAD{twelve}% 
                           \def\vsHEAD{nine}% 
                           \ifxxpt\else\xdef\f@ntsize{\HEAD}% 
                           \def\m@g{4}\def\s@ze{20.74}% 
                           \loadheadfonts\xxpttrue\fi 
                           \ifxiipt\else\xdef\f@ntsize{\smHEAD}% 
                           \def\m@g{1}\def\s@ze{12}% 
                           \loadxiiptfonts\xiipttrue\fi 
                           \ifixpt\else\xdef\f@ntsize{\vsHEAD}% 
                           \def\s@ze{9}% 
                           \loadsmallfonts\ixpttrue\fi 
                      \else 
                \ifnum#1=17\def\HEAD{seventeen}% 
                           \def\smHEAD{eleven}% 
                           \def\vsHEAD{eight}% 
                           \ifxviipt\else\xdef\f@ntsize{\HEAD}% 
                           \def\m@g{3}\def\s@ze{17.28}% 
                           \loadheadfonts\xviipttrue\fi 
                           \ifxipt\else\xdef\f@ntsize{\smHEAD}% 
                           \loadxiptfonts\xipttrue\fi 
                           \ifviiipt\else\xdef\f@ntsize{\vsHEAD}% 
                           \def\s@ze{8}% 
                           \loadsmallfonts\viiipttrue\fi 
                      \else\def\HEAD{fourteen}% 
                           \def\smHEAD{ten}% 
                           \def\vsHEAD{seven}% 
                           \ifxivpt\else\xdef\f@ntsize{\HEAD}% 
                           \def\m@g{2}\def\s@ze{14.4}% 
                           \loadheadfonts\xivpttrue\fi 
                           \ifxpt\else\xdef\f@ntsize{\smHEAD}% 
                           \def\s@ze{10}% 
                           \loadxptfonts\xpttrue\fi 
                           \ifviipt\else\xdef\f@ntsize{\vsHEAD}% 
                           \def\s@ze{7}% 
                           \loadviiptfonts\viipttrue\fi 
                \ifnum#1=14\else 
                \message{Header size should be 20, 17 or 14 point 
                              will now default to 14pt}\fi 
                \fi\fi\headfonts} 
% 
% Text in 12pt, 11pt or 10pt  
% 
\def\textsize#1#2{\def\textb@seline{#2}% 
                 \ifnum#1=12\def\TEXT{twelve}% 
                           \def\smTEXT{eight}% 
                           \def\vsTEXT{six}% 
                           \ifxiipt\else\xdef\f@ntsize{\TEXT}% 
                           \def\m@g{1}\def\s@ze{12}% 
                           \loadxiiptfonts\xiipttrue\fi 
                           \ifviiipt\else\xdef\f@ntsize{\smTEXT}% 
                           \def\s@ze{8}% 
                           \loadsmallfonts\viiipttrue\fi 
                           \ifvipt\else\xdef\f@ntsize{\vsTEXT}% 
                           \def\s@ze{6}% 
                           \loadviptfonts\vipttrue\fi 
                      \else 
                \ifnum#1=11\def\TEXT{eleven}% 
                           \def\smTEXT{seven}% 
                           \def\vsTEXT{five}% 
                           \ifxipt\else\xdef\f@ntsize{\TEXT}% 
                           \def\s@ze{11}% 
                           \loadxiptfonts\xipttrue\fi 
                           \ifviipt\else\xdef\f@ntsize{\smTEXT}% 
                           \loadviiptfonts\viipttrue\fi 
                           \ifvpt\else\xdef\f@ntsize{\vsTEXT}% 
                           \def\s@ze{5}% 
                           \loadvptfonts\vpttrue\fi 
                      \else\def\TEXT{ten}% 
                           \def\smTEXT{seven}% 
                           \def\vsTEXT{five}% 
                           \ifxpt\else\xdef\f@ntsize{\TEXT}% 
                           \loadxptfonts\xpttrue\fi 
                           \ifviipt\else\xdef\f@ntsize{\smTEXT}% 
                           \def\s@ze{7}% 
                           \loadviiptfonts\viipttrue\fi 
                           \ifvpt\else\xdef\f@ntsize{\vsTEXT}% 
                           \def\s@ze{5}% 
                           \loadvptfonts\vpttrue\fi 
                \ifnum#1=10\else 
                \message{Text size should be 12, 11 or 10 point 
                              will now default to 10pt}\fi 
                \fi\fi\textfonts} 
% 
% Small sized material in 10pt, 9pt or 8pt 
% 
\def\smallsize#1#2{\def\smallb@seline{#2}% 
                 \ifnum#1=10\def\SMALL{ten}% 
                           \def\smSMALL{seven}% 
                           \def\vsSMALL{five}% 
                           \ifxpt\else\xdef\f@ntsize{\SMALL}% 
                           \loadxptfonts\xpttrue\fi 
                           \ifviipt\else\xdef\f@ntsize{\smSMALL}% 
                           \def\s@ze{7}% 
                           \loadviiptfonts\viipttrue\fi 
                           \ifvpt\else\xdef\f@ntsize{\vsSMALL}% 
                           \def\s@ze{5}% 
                           \loadvptfonts\vpttrue\fi 
                       \else 
                 \ifnum#1=9\def\SMALL{nine}% 
                           \def\smSMALL{six}% 
                           \def\vsSMALL{five}% 
                           \ifixpt\else\xdef\f@ntsize{\SMALL}% 
                           \def\s@ze{9}% 
                           \loadsmallfonts\ixpttrue\fi 
                           \ifvipt\else\xdef\f@ntsize{\smSMALL}% 
                           \def\s@ze{6}% 
                           \loadviptfonts\vipttrue\fi 
                           \ifvpt\else\xdef\f@ntsize{\vsSMALL}% 
                           \def\s@ze{5}% 
                           \loadvptfonts\vpttrue\fi 
                       \else 
                           \def\SMALL{eight}% 
                           \def\smSMALL{six}% 
                           \def\vsSMALL{five}% 
                           \ifviiipt\else\xdef\f@ntsize{\SMALL}% 
                           \def\s@ze{8}% 
                           \loadsmallfonts\viiipttrue\fi 
                           \ifvipt\else\xdef\f@ntsize{\smSMALL}% 
                           \def\s@ze{6}% 
                           \loadviptfonts\vipttrue\fi 
                           \ifvpt\else\xdef\f@ntsize{\vsSMALL}% 
                           \def\s@ze{5}% 
                           \loadvptfonts\vpttrue\fi 
                 \ifnum#1=8\else\message{Small size should be 10, 9 or  
                            8 point will now default to 8pt}\fi 
                \fi\fi\smallfonts} 
\def\F@nt{\expandafter\font\csname} 
\def\Sk@w{\expandafter\skewchar\csname} 
\def\@nd{\endcsname} 
\def\@step#1{ scaled \magstep#1} 
\def\@half{ scaled \magstephalf} 
\def\@t#1{ at #1pt} 
% 
% For 14, 17 and 20 point fonts use \loadheadfonts 
% 
\def\loadheadfonts{\bigf@nts 
\F@nt \f@ntsize bdi\@nd=cmmib10 \@t{\s@ze}% 
\Sk@w \f@ntsize bdi\@nd='177 
\F@nt \f@ntsize bsy\@nd=cmbsy10 \@t{\s@ze}% 
\Sk@w \f@ntsize bsy\@nd='60 
\F@nt \f@ntsize bss\@nd=cmssbx10 \@t{\s@ze}} 
% 
% For 12 point fonts use \loadxiiptfonts 
% 
\def\loadxiiptfonts{\bigf@nts 
\F@nt \f@ntsize bdi\@nd=cmmib10 \@step{\m@g}% 
\Sk@w \f@ntsize bdi\@nd='177 
\F@nt \f@ntsize bsy\@nd=cmbsy10 \@step{\m@g}% 
\Sk@w \f@ntsize bsy\@nd='60 
\F@nt \f@ntsize bss\@nd=cmssbx10 \@step{\m@g}} 
% 
% For 11 point fonts use \loadxiptfonts 
% 
\def\loadxiptfonts{% 
\font\elevenrm=cmr10 \@half 
\font\eleveni=cmmi10 \@half 
\skewchar\eleveni='177 
\font\elevensy=cmsy10 \@half 
\skewchar\elevensy='60 
\font\elevenex=cmex10 \@half 
\font\elevenit=cmti10 \@half 
\font\elevensl=cmsl10 \@half 
\font\elevenbf=cmbx10 \@half 
\font\eleventt=cmtt10 \@half 
\ifams\font\elevenmsa=msam10 \@half 
\font\elevenmsb=msbm10 \@half\else\fi 
\font\elevenbdi=cmmib10 \@half 
\skewchar\elevenbdi='177 
\font\elevenbsy=cmbsy10 \@half 
\skewchar\elevenbsy='60 
\font\elevenbss=cmssbx10 \@half} 
% 
% For 10 point fonts use \loadxptfonts 
% 
\def\loadxptfonts{% 
\font\tenbdi=cmmib10 
\skewchar\tenbdi='177 
\font\tenbsy=cmbsy10  
\skewchar\tenbsy='60 
\ifams\font\tenmsa=msam10  
\font\tenmsb=msbm10\else\fi 
\font\tenbss=cmssbx10}%  
% 
% For 8 and 9 point fonts use \loadsmallfonts 
% 
\def\loadsmallfonts{\smallf@nts 
\ifams 
\F@nt \f@ntsize ex\@nd=cmex\s@ze 
\else 
\F@nt \f@ntsize ex\@nd=cmex10\fi 
\F@nt \f@ntsize it\@nd=cmti\s@ze 
\F@nt \f@ntsize sl\@nd=cmsl\s@ze 
\F@nt \f@ntsize tt\@nd=cmtt\s@ze} 
% 
% For 7 point fonts use \loadviiptfonts 
% 
\def\loadviiptfonts{% 
\font\sevenit=cmti7 
\font\sevensl=cmsl8 at 7pt 
\ifams\font\sevenmsa=msam7  
\font\sevenmsb=msbm7 
\font\sevenex=cmex7 
\font\sevenbsy=cmbsy7 
\font\sevenbdi=cmmib7\else 
\font\sevenex=cmex10 
\font\sevenbsy=cmbsy10 at 7pt 
\font\sevenbdi=cmmib10 at 7pt\fi 
\skewchar\sevenbsy='60 
\skewchar\sevenbdi='177 
\font\sevenbss=cmssbx10 at 7pt}%  
% 
%  For 6 point fonts use \loadviptfonts 
% 
\def\loadviptfonts{\smallf@nts 
\ifams\font\sixex=cmex7 at 6pt\else 
\font\sixex=cmex10\fi 
\font\sixit=cmti7 at 6pt} 
% 
% For 5 point fonts use \loadvptfonts 
% 
\def\loadvptfonts{% 
\font\fiveit=cmti7 at 5pt 
\ifams\font\fiveex=cmex7 at 5pt 
\font\fivebdi=cmmib5 
\font\fivebsy=cmbsy5 
\font\fivemsa=msam5  
\font\fivemsb=msbm5\else 
\font\fiveex=cmex10 
\font\fivebdi=cmmib10 at 5pt 
\font\fivebsy=cmbsy10 at 5pt\fi 
\skewchar\fivebdi='177 
\skewchar\fivebsy='60 
\font\fivebss=cmssbx10 at 5pt} 
\def\bigf@nts{% 
\F@nt \f@ntsize rm\@nd=cmr10 \@step{\m@g}% 
\F@nt \f@ntsize i\@nd=cmmi10 \@step{\m@g}% 
\Sk@w \f@ntsize i\@nd='177 
\F@nt \f@ntsize sy\@nd=cmsy10 \@step{\m@g}% 
\Sk@w \f@ntsize sy\@nd='60 
\F@nt \f@ntsize ex\@nd=cmex10 \@step{\m@g}% 
\F@nt \f@ntsize it\@nd=cmti10 \@step{\m@g}% 
\F@nt \f@ntsize sl\@nd=cmsl10 \@step{\m@g}% 
\F@nt \f@ntsize bf\@nd=cmbx10 \@step{\m@g}% 
\F@nt \f@ntsize tt\@nd=cmtt10 \@step{\m@g}% 
\ifams 
\F@nt \f@ntsize msa\@nd=msam10 \@step{\m@g}% 
\F@nt \f@ntsize msb\@nd=msbm10 \@step{\m@g}\else\fi} 
\def\smallf@nts{% 
\F@nt \f@ntsize rm\@nd=cmr\s@ze 
\F@nt \f@ntsize i\@nd=cmmi\s@ze  
\Sk@w \f@ntsize i\@nd='177 
\F@nt \f@ntsize sy\@nd=cmsy\s@ze 
\Sk@w \f@ntsize sy\@nd='60 
\F@nt \f@ntsize bf\@nd=cmbx\s@ze  
\ifams 
\F@nt \f@ntsize bdi\@nd=cmmib\s@ze  
\F@nt \f@ntsize bsy\@nd=cmbsy\s@ze  
\F@nt \f@ntsize msa\@nd=msam\s@ze  
\F@nt \f@ntsize msb\@nd=msbm\s@ze 
\else 
\F@nt \f@ntsize bdi\@nd=cmmib10 \@t{\s@ze}%  
\F@nt \f@ntsize bsy\@nd=cmbsy10 \@t{\s@ze}\fi  
\Sk@w \f@ntsize bdi\@nd='177 
\Sk@w \f@ntsize bsy\@nd='60 
\F@nt \f@ntsize bss\@nd=cmssbx10 \@t{\s@ze}}%  
% 
% Fonts for headings  
% 
\def\headfonts{% 
\textfont0=\csname\HEAD rm\@nd         
\scriptfont0=\csname\smHEAD rm\@nd 
\scriptscriptfont0=\csname\vsHEAD rm\@nd 
\def\rm{\fam0\csname\HEAD rm\@nd 
\def\sc{\csname\smHEAD rm\@nd}}% 
\textfont1=\csname\HEAD i\@nd          
\scriptfont1=\csname\smHEAD i\@nd 
\scriptscriptfont1=\csname\vsHEAD i\@nd 
\textfont2=\csname\HEAD sy\@nd         
\scriptfont2=\csname\smHEAD sy\@nd 
\scriptscriptfont2=\csname\vsHEAD sy\@nd 
\textfont3=\csname\HEAD ex\@nd         
\scriptfont3=\csname\smHEAD ex\@nd 
\scriptscriptfont3=\csname\smHEAD ex\@nd 
\textfont\itfam=\csname\HEAD it\@nd    
\scriptfont\itfam=\csname\smHEAD it\@nd 
\scriptscriptfont\itfam=\csname\vsHEAD it\@nd 
\def\it{\fam\itfam\csname\HEAD it\@nd 
\def\sc{\csname\smHEAD it\@nd}}% 
\textfont\slfam=\csname\HEAD sl\@nd    
\def\sl{\fam\slfam\csname\HEAD sl\@nd 
\def\sc{\csname\smHEAD sl\@nd}}% 
\textfont\bffam=\csname\HEAD bf\@nd    
\scriptfont\bffam=\csname\smHEAD bf\@nd 
\scriptscriptfont\bffam=\csname\vsHEAD bf\@nd 
\def\bf{\fam\bffam\csname\HEAD bf\@nd 
\def\sc{\csname\smHEAD bf\@nd}}% 
\textfont\ttfam=\csname\HEAD tt\@nd    
\def\tt{\fam\ttfam\csname\HEAD tt\@nd}% 
\textfont\bdifam=\csname\HEAD bdi\@nd  
\scriptfont\bdifam=\csname\smHEAD bdi\@nd 
\scriptscriptfont\bdifam=\csname\vsHEAD bdi\@nd 
\def\bdi{\fam\bdifam\csname\HEAD bdi\@nd}% 
\textfont\bsyfam=\csname\HEAD bsy\@nd  
\scriptfont\bsyfam=\csname\smHEAD bsy\@nd 
\def\bsy{\fam\bsyfam\csname\HEAD bsy\@nd}% 
\textfont\bssfam=\csname\HEAD bss\@nd  
\scriptfont\bssfam=\csname\smHEAD bss\@nd 
\scriptscriptfont\bssfam=\csname\vsHEAD bss\@nd 
\def\bss{\fam\bssfam\csname\HEAD bss\@nd}% 
\ifams 
\textfont\msafam=\csname\HEAD msa\@nd  
\scriptfont\msafam=\csname\smHEAD msa\@nd 
\scriptscriptfont\msafam=\csname\vsHEAD msa\@nd 
\textfont\msbfam=\csname\HEAD msb\@nd  
\scriptfont\msbfam=\csname\smHEAD msb\@nd 
\scriptscriptfont\msbfam=\csname\vsHEAD msb\@nd 
\else\fi 
\normalbaselineskip=\headb@seline pt% 
\setbox\strutbox=\hbox{\vrule height.7\normalbaselineskip  
depth.3\baselineskip width0pt}% 
\def\sc{\csname\smHEAD rm\@nd}\normalbaselines\bf} 
% 
% Fonts for text 
% 
\def\textfonts{% 
\textfont0=\csname\TEXT rm\@nd         
\scriptfont0=\csname\smTEXT rm\@nd 
\scriptscriptfont0=\csname\vsTEXT rm\@nd 
\def\rm{\fam0\csname\TEXT rm\@nd 
\def\sc{\csname\smTEXT rm\@nd}}% 
\textfont1=\csname\TEXT i\@nd          
\scriptfont1=\csname\smTEXT i\@nd 
\scriptscriptfont1=\csname\vsTEXT i\@nd 
\textfont2=\csname\TEXT sy\@nd         
\scriptfont2=\csname\smTEXT sy\@nd 
\scriptscriptfont2=\csname\vsTEXT sy\@nd 
\textfont3=\csname\TEXT ex\@nd         
\scriptfont3=\csname\smTEXT ex\@nd 
\scriptscriptfont3=\csname\smTEXT ex\@nd 
\textfont\itfam=\csname\TEXT it\@nd    
\scriptfont\itfam=\csname\smTEXT it\@nd 
\scriptscriptfont\itfam=\csname\vsTEXT it\@nd 
\def\it{\fam\itfam\csname\TEXT it\@nd 
\def\sc{\csname\smTEXT it\@nd}}% 
\textfont\slfam=\csname\TEXT sl\@nd    
\def\sl{\fam\slfam\csname\TEXT sl\@nd 
\def\sc{\csname\smTEXT sl\@nd}}% 
\textfont\bffam=\csname\TEXT bf\@nd    
\scriptfont\bffam=\csname\smTEXT bf\@nd 
\scriptscriptfont\bffam=\csname\vsTEXT bf\@nd 
\def\bf{\fam\bffam\csname\TEXT bf\@nd 
\def\sc{\csname\smTEXT bf\@nd}}% 
\textfont\ttfam=\csname\TEXT tt\@nd    
\def\tt{\fam\ttfam\csname\TEXT tt\@nd}% 
\textfont\bdifam=\csname\TEXT bdi\@nd  
\scriptfont\bdifam=\csname\smTEXT bdi\@nd 
\scriptscriptfont\bdifam=\csname\vsTEXT bdi\@nd 
\def\bdi{\fam\bdifam\csname\TEXT bdi\@nd}% 
\textfont\bsyfam=\csname\TEXT bsy\@nd  
\scriptfont\bsyfam=\csname\smTEXT bsy\@nd 
\def\bsy{\fam\bsyfam\csname\TEXT bsy\@nd}% 
\textfont\bssfam=\csname\TEXT bss\@nd  
\scriptfont\bssfam=\csname\smTEXT bss\@nd 
\scriptscriptfont\bssfam=\csname\vsTEXT bss\@nd 
\def\bss{\fam\bssfam\csname\TEXT bss\@nd}% 
\ifams 
\textfont\msafam=\csname\TEXT msa\@nd  
\scriptfont\msafam=\csname\smTEXT msa\@nd 
\scriptscriptfont\msafam=\csname\vsTEXT msa\@nd 
\textfont\msbfam=\csname\TEXT msb\@nd  
\scriptfont\msbfam=\csname\smTEXT msb\@nd 
\scriptscriptfont\msbfam=\csname\vsTEXT msb\@nd 
\else\fi 
\normalbaselineskip=\textb@seline pt 
\setbox\strutbox=\hbox{\vrule height.7\normalbaselineskip  
depth.3\baselineskip width0pt}% 
\everymath{}% 
\def\sc{\csname\smTEXT rm\@nd}\normalbaselines\rm} 
% 
% Fonts for small material (captions, footnotes etc) 
% 
\def\smallfonts{% 
\textfont0=\csname\SMALL rm\@nd         
\scriptfont0=\csname\smSMALL rm\@nd 
\scriptscriptfont0=\csname\vsSMALL rm\@nd 
\def\rm{\fam0\csname\SMALL rm\@nd 
\def\sc{\csname\smSMALL rm\@nd}}% 
\textfont1=\csname\SMALL i\@nd          
\scriptfont1=\csname\smSMALL i\@nd 
\scriptscriptfont1=\csname\vsSMALL i\@nd 
\textfont2=\csname\SMALL sy\@nd         
\scriptfont2=\csname\smSMALL sy\@nd 
\scriptscriptfont2=\csname\vsSMALL sy\@nd 
\textfont3=\csname\SMALL ex\@nd         
\scriptfont3=\csname\smSMALL ex\@nd 
\scriptscriptfont3=\csname\smSMALL ex\@nd 
\textfont\itfam=\csname\SMALL it\@nd    
\scriptfont\itfam=\csname\smSMALL it\@nd 
\scriptscriptfont\itfam=\csname\vsSMALL it\@nd 
\def\it{\fam\itfam\csname\SMALL it\@nd 
\def\sc{\csname\smSMALL it\@nd}}% 
\textfont\slfam=\csname\SMALL sl\@nd    
\def\sl{\fam\slfam\csname\SMALL sl\@nd 
\def\sc{\csname\smSMALL sl\@nd}}% 
\textfont\bffam=\csname\SMALL bf\@nd    
\scriptfont\bffam=\csname\smSMALL bf\@nd 
\scriptscriptfont\bffam=\csname\vsSMALL bf\@nd 
\def\bf{\fam\bffam\csname\SMALL bf\@nd 
\def\sc{\csname\smSMALL bf\@nd}}% 
\textfont\ttfam=\csname\SMALL tt\@nd    
\def\tt{\fam\ttfam\csname\SMALL tt\@nd}% 
\textfont\bdifam=\csname\SMALL bdi\@nd  
\scriptfont\bdifam=\csname\smSMALL bdi\@nd 
\scriptscriptfont\bdifam=\csname\vsSMALL bdi\@nd 
\def\bdi{\fam\bdifam\csname\SMALL bdi\@nd}% 
\textfont\bsyfam=\csname\SMALL bsy\@nd  
\scriptfont\bsyfam=\csname\smSMALL bsy\@nd 
\def\bsy{\fam\bsyfam\csname\SMALL bsy\@nd}% 
\textfont\bssfam=\csname\SMALL bss\@nd  
\scriptfont\bssfam=\csname\smSMALL bss\@nd 
\scriptscriptfont\bssfam=\csname\vsSMALL bss\@nd 
\def\bss{\fam\bssfam\csname\SMALL bss\@nd}% 
\ifams 
\textfont\msafam=\csname\SMALL msa\@nd  
\scriptfont\msafam=\csname\smSMALL msa\@nd 
\scriptscriptfont\msafam=\csname\vsSMALL msa\@nd 
\textfont\msbfam=\csname\SMALL msb\@nd  
\scriptfont\msbfam=\csname\smSMALL msb\@nd 
\scriptscriptfont\msbfam=\csname\vsSMALL msb\@nd 
\else\fi 
\normalbaselineskip=\smallb@seline pt% 
\setbox\strutbox=\hbox{\vrule height.7\normalbaselineskip  
depth.3\baselineskip width0pt}% 
\everymath{}% 
\def\sc{\csname\smSMALL rm\@nd}\normalbaselines\rm}% 
\everydisplay{\indenteddisplay 
   \gdef\labeltype{\eqlabel}}% 
% 
%%%%%%%%%%%%%%%%%%%%%%%%%%%%%%%%%%%%%%%%%%%%%%%%%%%%%%%%%%% 
%                                                         % 
%  Macros to define extra maths symbols                   % 
%                                                         % 
%%%%%%%%%%%%%%%%%%%%%%%%%%%%%%%%%%%%%%%%%%%%%%%%%%%%%%%%%%% 
% 
\def\hexnumber@#1{\ifcase#1 0\or 1\or 2\or 3\or 4\or 5\or 6\or 7\or 8\or 
 9\or A\or B\or C\or D\or E\or F\fi} 
\edef\bffam@{\hexnumber@\bffam} 
\edef\bdifam@{\hexnumber@\bdifam} 
\edef\bsyfam@{\hexnumber@\bsyfam} 
\def\undefine#1{\let#1\undefined} 
\def\newsymbol#1#2#3#4#5{\let\next@\relax 
 \ifnum#2=\thr@@\let\next@\bdifam@\else 
 \ifams 
 \ifnum#2=\@ne\let\next@\msafam@\else 
 \ifnum#2=\tw@\let\next@\msbfam@\fi\fi 
 \fi\fi 
 \mathchardef#1="#3\next@#4#5} 
\def\mathhexbox@#1#2#3{\relax 
 \ifmmode\mathpalette{}{\m@th\mathchar"#1#2#3}% 
 \else\leavevmode\hbox{$\m@th\mathchar"#1#2#3$}\fi} 

\def\bi#1{{\fam\bdifam\relax#1}} 
% 
% If file amsmacro is not in current directory 
% or somewhere with set path add path before 
% file name in following line 
% 
\ifams\input amsmacro\fi 
% 
% Bold italic Greek characters 
% 
\newsymbol\bitGamma 3000 
\newsymbol\bitDelta 3001 
\newsymbol\bitTheta 3002 
\newsymbol\bitLambda 3003 
\newsymbol\bitXi 3004 
\newsymbol\bitPi 3005 
\newsymbol\bitSigma 3006 
\newsymbol\bitUpsilon 3007 
\newsymbol\bitPhi 3008 
\newsymbol\bitPsi 3009 
\newsymbol\bitOmega 300A 
\newsymbol\balpha 300B 
\newsymbol\bbeta 300C 
\newsymbol\bgamma 300D 
\newsymbol\bdelta 300E 
\newsymbol\bepsilon 300F 
\newsymbol\bzeta 3010 
\newsymbol\bfeta 3011 
\newsymbol\btheta 3012 
\newsymbol\biota 3013 
\newsymbol\bkappa 3014 
\newsymbol\blambda 3015 
\newsymbol\bmu 3016 
\newsymbol\bnu 3017 
\newsymbol\bxi 3018 
\newsymbol\bpi 3019 
\newsymbol\brho 301A 
\newsymbol\bsigma 301B 
\newsymbol\btau 301C 
\newsymbol\bupsilon 301D 
\newsymbol\bphi 301E 
\newsymbol\bchi 301F 
\newsymbol\bpsi 3020 
\newsymbol\bomega 3021 
\newsymbol\bvarepsilon 3022 
\newsymbol\bvartheta 3023 
\newsymbol\bvaromega 3024 
\newsymbol\bvarrho 3025 
\newsymbol\bvarzeta 3026 
\newsymbol\bvarphi 3027 
\newsymbol\bpartial 3040 
\newsymbol\bell 3060 
\newsymbol\bimath 307B 
\newsymbol\bjmath 307C 
\mathchardef\binfty "0\bsyfam@31 
\mathchardef\bnabla "0\bsyfam@72 
\mathchardef\bdot "2\bsyfam@01 
\mathchardef\bGamma "0\bffam@00 
\mathchardef\bDelta "0\bffam@01 
\mathchardef\bTheta "0\bffam@02 
\mathchardef\bLambda "0\bffam@03 
\mathchardef\bXi "0\bffam@04 
\mathchardef\bPi "0\bffam@05 
\mathchardef\bSigma "0\bffam@06 
\mathchardef\bUpsilon "0\bffam@07 
\mathchardef\bPhi "0\bffam@08 
\mathchardef\bPsi "0\bffam@09 
\mathchardef\bOmega "0\bffam@0A 
\mathchardef\itGamma "0100 
\mathchardef\itDelta "0101 
\mathchardef\itTheta "0102 
\mathchardef\itLambda "0103 
\mathchardef\itXi "0104 
\mathchardef\itPi "0105 
\mathchardef\itSigma "0106 
\mathchardef\itUpsilon "0107 
\mathchardef\itPhi "0108 
\mathchardef\itPsi "0109 
\mathchardef\itOmega "010A 
\mathchardef\Gamma "0000 
\mathchardef\Delta "0001 
\mathchardef\Theta "0002 
\mathchardef\Lambda "0003 
\mathchardef\Xi "0004 
\mathchardef\Pi "0005 
\mathchardef\Sigma "0006 
\mathchardef\Upsilon "0007 
\mathchardef\Phi "0008 
\mathchardef\Psi "0009 
\mathchardef\Omega "000A 
% 
% Counter definitions 
% 
\newcount\firstpage  \firstpage=1  % start page no 
\newcount\jnl                      % journal no 
\newcount\secno                    % section number 
\newcount\subno                    % number of subsection 
\newcount\subsubno                 % number of subsubsection 
\newcount\appno                    % appendix number 
\newcount\tabno                    % table number 
\newcount\figno                    % figure number 
\newcount\countno                  % equation numbers 
\newcount\refno                    % reference number 
\newcount\eqlett     \eqlett=97    % equation letter 
\newif\ifletter 
\newif\ifwide 
\newif\ifnotfull 
\newif\ifaligned 
\newif\ifnumbysec   
\newif\ifappendix 
\newif\ifnumapp 
\newif\ifssf 
\newif\ifppt 
\newdimen\t@bwidth 
\newdimen\c@pwidth 
\newdimen\digitwidth                    %character width 
\newdimen\argwidth                      %argument width 
\newdimen\secindent    \secindent=5pc   %indentation of maths  
\newdimen\textind    \textind=16pt      %indentation of text 
\newdimen\tempval                       %temporary value 
\newskip\beforesecskip 
\def\beforesecspace{\vskip\beforesecskip\relax} 
\newskip\beforesubskip 
\def\beforesubspace{\vskip\beforesubskip\relax} 
\newskip\beforesubsubskip 
\def\beforesubsubspace{\vskip\beforesubsubskip\relax} 
\newskip\secskip 
\def\secspace{\vskip\secskip\relax} 
\newskip\subskip 
\def\subspace{\vskip\subskip\relax} 
\newskip\insertskip 
\def\insertspace{\vskip\insertskip\relax} 
\def\sp@ce{\ifx\next*\let\next=\@ssf 
               \else\let\next=\@nossf\fi\next} 
\def\@ssf#1{\nobreak\secspace\global\ssftrue\nobreak} 
\def\@nossf{\nobreak\secspace\nobreak\noindent\ignorespaces} 
\def\subsp@ce{\ifx\next*\let\next=\@sssf 
               \else\let\next=\@nosssf\fi\next} 
\def\@sssf#1{\nobreak\subspace\global\ssftrue\nobreak} 
\def\@nosssf{\nobreak\subspace\nobreak\noindent\ignorespaces} 
\beforesecskip=24pt plus12pt minus8pt 
\beforesubskip=12pt plus6pt minus4pt 
\beforesubsubskip=12pt plus6pt minus4pt 
\secskip=12pt plus 2pt minus 2pt 
\subskip=6pt plus3pt minus2pt 
\insertskip=18pt plus6pt minus6pt% 
\fontdimen16\tensy=2.7pt 
\fontdimen17\tensy=2.7pt 
% 
% Labels etc for cross referencing macros 
% 
\def\eqlabel{(\ifappendix\applett 
               \ifnumbysec\ifnum\secno>0 \the\secno\fi.\fi 
               \else\ifnumbysec\the\secno.\fi\fi\the\countno)} 
\def\seclabel{\ifappendix\ifnumapp\else\applett\fi 
    \ifnum\secno>0 \the\secno 
    \ifnumbysec\ifnum\subno>0.\the\subno\fi\fi\fi 
    \else\the\secno\fi\ifnum\subno>0.\the\subno 
         \ifnum\subsubno>0.\the\subsubno\fi\fi} 
\def\tablabel{\ifappendix\applett\fi\the\tabno} 
\def\figlabel{\ifappendix\applett\fi\the\figno} 
\def\gac{\global\advance\countno by 1} 
% 
% Redefinition of footnote macros to lose rule and remove indentation 
% 
 
\def\vfootnote#1{\insert\footins\bgroup 
\interlinepenalty=\interfootnotelinepenalty 
\splittopskip=\ht\strutbox % top baseline for broken footnotes 
\splitmaxdepth=\dp\strutbox \floatingpenalty=20000 
\leftskip=0pt \rightskip=0pt \spaceskip=0pt \xspaceskip=0pt% 
\noindent\smallfonts\rm #1\ \ignorespaces\footstrut\futurelet\next\fo@t} 
% 
% Redefinition of endinsert to give more controllable 
% space around  tables and figures 
% 
\def\endinsert{\egroup 
    \if@mid \dimen@=\ht0 \advance\dimen@ by\dp0 
       \advance\dimen@ by12\p@ \advance\dimen@ by\pagetotal 
       \ifdim\dimen@>\pagegoal \@midfalse\p@gefalse\fi\fi 
    \if@mid \insertspace \box0 \par \ifdim\lastskip<\insertskip 
    \removelastskip \penalty-200 \insertspace \fi 
    \else\insert\topins{\penalty100 
       \splittopskip=0pt \splitmaxdepth=\maxdimen  
       \floatingpenalty=0 
       \ifp@ge \dimen@=\dp0 
       \vbox to\vsize{\unvbox0 \kern-\dimen@}% 
       \else\box0\nobreak\insertspace\fi}\fi\endgroup}    
% 
% special macros for display equations 
% 
% for indentation of turned over lines in mathematics 
% 
\def\ind{\hbox to \secindent{\hfill}} 
% 
% for turned over equals sign to left of maths indent 
% 

% 
% for other signs to left of maths indent 
% 
 
% 
% displayed equation indented  
% 
\def\indeqn#1{\alignedfalse\displ@y\halign{\hbox to \displaywidth 
    {$\ind\@lign\displaystyle##\hfil$}\crcr #1\crcr}} 
% 
% displayed equation indented with alignments 
% 
\def\indalign#1{\alignedtrue\displ@y \tabskip=0pt  
  \halign to\displaywidth{\ind$\@lign\displaystyle{##}$\tabskip=0pt 
    &$\@lign\displaystyle{{}##}$\hfill\tabskip=\centering 
    &\llap{$\@lign\hbox{\rm##}$}\tabskip=0pt\crcr 
    #1\crcr}} 
\def\indenteddisplay#1$${\indispl@y{#1 }} 
\def\indispl@y#1{\disptest#1\eqalignno\eqalignno\disptest} 
\def\disptest#1\eqalignno#2\eqalignno#3\disptest{% 
    \ifx#3\eqalignno 
    \indalign#2% 
    \else\indeqn{#1}\fi$$} 
% 
% Roman small caps (if in Roman \sc gives small caps) 
% 
 
% 
% Italic small caps (if in italic \sc gives italic small caps) 
% 
 
% 
% Bold small caps (if in bold \sc gives bold small caps) 
% 
 
% 
% Small caps in maths 
% 
 
% 
% Miscellaneous definitions 
% 

\def\ns{\noalign{\vskip-3pt}}

% 
 
% 
% Bold h bar 
% 
\def\bhbar{\rlap{\kern1pt\raise.4ex\hbox{\bf\char'40}}\bi{h}} 

\def\e{{\rm e}} 
 
\def\frac#1#2{{#1\over#2}} 
\ifams 
\def\lap{\lesssim} 
\def\gap{\gtrsim}

\let\leq=\leqslant

\let\geq=\geqslant 
\else

\def\gap{\;\lower3pt\hbox{$\buildrel > \over \sim$}\;}% 
\def\lap{\;\lower3pt\hbox{$\buildrel < \over \sim$}\;}\fi 
 
\chardef\ii="10 
\def\tqs{\hbox to 25pt{\hfil}}

\def\Bbbone{1\kern-.22em {\rm l}} 
% 
% Primes to display summations and products  
% which also have sub or superscripts 
% 
\def\rp{\raise8pt\hbox{$\scriptstyle\prime$}} 
% 
% then use \sum^{...}_{...}\rp or \prod^{...}_{...}\rp. 
% 
% Shadow brackets 
% 
% Single brackets for normal size only 
% 

% 
% Variable size for display style 
% 
\def\[#1\]{\setbox0=\hbox{$\dsty#1$}\argwidth=\wd0 
    \setbox0=\hbox{$\left[\box0\right]$}\advance\argwidth by -\wd0 
    \left[\kern.3\argwidth\box0\kern.3\argwidth\right]} 
% 
% Variable size for text style 
% 
\def\lsb#1\rsb{\setbox0=\hbox{$#1$}\argwidth=\wd0 
    \setbox0=\hbox{$\left[\box0\right]$}\advance\argwidth by -\wd0 
    \left[\kern.3\argwidth\box0\kern.3\argwidth\right]} 
% 
 
% 
% Square for end of theorems 
% 
 
% 
\def\pt(#1){({\it #1\/})} 
\let\dsty=\displaystyle

% 
% Definition for Nuclear Physics Keyword abstract 
% 
\def\reactions#1{\vskip 12pt plus2pt minus2pt%              
\vbox{\hbox{\kern\secindent\vrule\kern12pt% 
\vbox{\kern0.5pt\vbox{\hsize=24pc\parindent=0pt\smallfonts\rm NUCLEAR  
REACTIONS\strut\quad #1\strut}\kern0.5pt}\kern12pt\vrule}}} 
% 
% Definition for slashed characters 
% 
\def\slashchar#1{\setbox0=\hbox{$#1$}\dimen0=\wd0% 
\setbox1=\hbox{/}\dimen1=\wd1% 
\ifdim\dimen0>\dimen1%                         
\rlap{\hbox to \dimen0{\hfil/\hfil}}#1\else                                         
\rlap{\hbox to \dimen1{\hfil$#1$\hfil}}/\fi} 
% 
% Redefine \textindent for use in \item 
% 
\def\textindent#1{\noindent\hbox to \parindent{#1\hss}\ignorespaces} 
% 
% Symbols and curves for use in figure captions 
% 
\def\opencirc{\raise1pt\hbox{$\scriptstyle{\bigcirc}$}} 
 
\ifams 
\def\opensqr{\hbox{$\square$}} 
 
\def\opentridown{\hbox{$\triangledown$}}

\else 
\def\opensqr{\vbox{\hrule height.4pt\hbox{\vrule width.4pt height3.5pt 
    \kern3.5pt\vrule width.4pt}\hrule height.4pt}} 
 
\def\opentridown{\raise1pt\hbox{$\scriptstyle\bigtriangledown$}}

           %  These produce the 
                   %  equivalent open character 
           %  to be filled in. 
\fi

% 
% Redefinition of \cases 
% 
\def\m@th{\mathsurround=0pt} 
% 
% Displaystyle now used for first term 
% 
\def\cases#1{% 
\left\{\,\vcenter{\normalbaselines\openup1\jot\m@th% 
     \ialign{$\displaystyle##\hfil$&\rm\tqs##\hfil\crcr#1\crcr}}\right.}% 
% 
% Original version of cases now called \oldcases 
% 
\def\oldcases#1{\left\{\,\vcenter{\normalbaselines\m@th 
    \ialign{$##\hfil$&\rm\quad##\hfil\crcr#1\crcr}}\right.} 
% 
% Cases with number at end each line (using automatic numbering) 
% 
\def\numcases#1{\left\{\,\vcenter{\baselineskip=15pt\m@th% 
     \ialign{$\displaystyle##\hfil$&\rm\tqs##\hfil 
     \crcr#1\crcr}}\right.\hfill 
     \vcenter{\baselineskip=15pt\m@th% 
     \ialign{\rlap{$\phantom{\displaystyle##\hfil}$}\tabskip=0pt&\en 
     \rlap{\phantom{##\hfil}}\crcr#1\crcr}}} 
\def\ptnumcases#1{\left\{\,\vcenter{\baselineskip=15pt\m@th% 
     \ialign{$\displaystyle##\hfil$&\rm\tqs##\hfil 
     \crcr#1\crcr}}\right.\hfill 
     \vcenter{\baselineskip=15pt\m@th% 
     \ialign{\rlap{$\phantom{\displaystyle##\hfil}$}\tabskip=0pt&\enpt 
     \rlap{\phantom{##\hfil}}\crcr#1\crcr}}\global\eqlett=97 
     \global\advance\countno by 1} 
% 
% for equation numbers instead of \eqno 
% 
\def\eq(#1){\ifaligned\@mp(#1)\else\hfill\llap{{\rm (#1)}}\fi} 
\def\ceq(#1){\ns\ns\ifaligned\@mp\fi\eq(#1)\cr\ns\ns} 
\def\eqpt(#1#2){\ifaligned\@mp(#1{\it #2\/}) 
                    \else\hfill\llap{{\rm (#1{\it #2\/})}}\fi} 
\let\eqno=\eq 
% 
% Automatic numbering of equations 
% 
\countno=1 
 
\def\aleq{&\rm(\ifappendix\applett 
               \ifnumbysec\ifnum\secno>0 \the\secno\fi.\fi 
               \else\ifnumbysec\the\secno.\fi\fi\the\countno} 
\def\noaleq{\hfill\llap\bgroup\rm(\ifappendix\applett 
               \ifnumbysec\ifnum\secno>0 \the\secno\fi.\fi 
               \else\ifnumbysec\the\secno.\fi\fi\the\countno} 
\def\@mp{&} 
\def\en{\ifaligned\aleq)\else\noaleq)\egroup\fi\gac} 
\def\cen{\ns\ns\ifaligned\@mp\fi\en\cr\ns\ns} 
\def\enpt{\ifaligned\aleq{\it\char\the\eqlett})\else 
    \noaleq{\it\char\the\eqlett})\egroup\fi 
    \global\advance\eqlett by 1} 
\def\endpt{\ifaligned\aleq{\it\char\the\eqlett})\else 
    \noaleq{\it\char\the\eqlett})\egroup\fi 
    \global\eqlett=97\gac} 
% 
% abbreviations for Institute of Physics Publishing journals 
% 

        %1968-87 
   %1988 and onwards 
     %1968--1988 
        %1989 and onwards 

           %1975--1988 
     %1989 and onwards 
 
                 %1990 and onwards 

% 
% Other commonly quoted journals 
% 

% 
\headline={\ifodd\pageno{\ifnum\pageno=\firstpage\hfill 
   \else\rrhead\fi}\else\lrhead\fi} 
\def\rrhead{\textfonts\hskip\secindent\it 
    \shorttitle\hfill\rm\folio} 
\def\lrhead{\textfonts\hbox to\secindent{\rm\folio\hss}% 
    \it\aunames\hss} 
\footline={\ifnum\pageno=\firstpage \hfill\textfonts\rm\folio\fi} 
\def\@rticle#1#2{\vglue.5pc 
    {\parindent=\secindent \bf #1\par} 
     \vskip2.5pc 
    {\exhyphenpenalty=10000\hyphenpenalty=10000 
     \baselineskip=18pt\raggedright\noindent 
     \headfonts\bf#2\par}\futurelet\next\sh@rttitle}% 
\def\title#1{\gdef\shorttitle{#1} 
    \vglue4pc{\exhyphenpenalty=10000\hyphenpenalty=10000  
    \baselineskip=18pt  
    \raggedright\parindent=0pt 
    \headfonts\bf#1\par}\futurelet\next\sh@rttitle}  

\def\article#1#2{\gdef\shorttitle{#2}\@rticle{#1}{#2}}  
\def\review#1{\gdef\shorttitle{#1}% 
    \@rticle{REVIEW \ifpbm\else ARTICLE\fi}{#1}} 
\def\topical#1{\gdef\shorttitle{#1}% 
    \@rticle{TOPICAL REVIEW}{#1}} 
\def\comment#1{\gdef\shorttitle{#1}% 
    \@rticle{COMMENT}{#1}} 
\def\note#1{\gdef\shorttitle{#1}% 
    \@rticle{NOTE}{#1}} 
\def\prelim#1{\gdef\shorttitle{#1}% 
    \@rticle{PRELIMINARY COMMUNICATION}{#1}} 
\def\letter#1{\gdef\shorttitle{Letter to the Editor}% 
     \gdef\aunames{Letter to the Editor} 
     \global\lettertrue\ifnum\jnl=7\global\letterfalse\fi 
     \@rticle{LETTER TO THE EDITOR}{#1}} 
\def\sh@rttitle{\ifx\next[\let\next=\sh@rt 
                \else\let\next=\f@ll\fi\next} 
\def\sh@rt[#1]{\gdef\shorttitle{#1}} 
\def\f@ll{} 
\def\author#1{\ifletter\else\gdef\aunames{#1}\fi\vskip1.5pc 
    {\parindent=\secindent   
     \hang\textfonts   
     \ifppt\bf\else\rm\fi#1\par}   
     \ifppt\bigskip\else\smallskip\fi 
     \futurelet\next\@unames} 
\def\@unames{\ifx\next[\let\next=\short@uthor 
                 \else\let\next=\@uthor\fi\next} 
\def\short@uthor[#1]{\gdef\aunames{#1}} 
\def\@uthor{} 
\def\address#1{{\parindent=\secindent 
    \exhyphenpenalty=10000\hyphenpenalty=10000 
\ifppt\textfonts\else\smallfonts\fi\hang\raggedright\rm#1\par}% 
    \ifppt\bigskip\fi} 
\def\jl#1{\global\jnl=#1} 
\jl{0}% 
\def\journal{\ifnum\jnl=1 J. Phys.\ A: Math.\ Gen.\  
        \else\ifnum\jnl=2 J. Phys.\ B: At.\ Mol.\ Opt.\ Phys.\  
        \else\ifnum\jnl=3 J. Phys.:\ Condens. Matter\  
        \else\ifnum\jnl=4 J. Phys.\ G: Nucl.\ Part.\ Phys.\  
        \else\ifnum\jnl=5 Inverse Problems\  
        \else\ifnum\jnl=6 Class. Quantum Grav.\  
        \else\ifnum\jnl=7 Network\  
        \else\ifnum\jnl=8 Nonlinearity\ 
        \else\ifnum\jnl=9 Quantum Opt.\ 
        \else\ifnum\jnl=10 Waves in Random Media\ 
        \else\ifnum\jnl=11 Pure Appl. Opt.\  
        \else\ifnum\jnl=12 Phys. Med. Biol.\ 
        \else\ifnum\jnl=13 Modelling Simulation Mater.\ Sci.\ Eng.\  
        \else\ifnum\jnl=14 Plasma Phys. Control. Fusion\  
        \else\ifnum\jnl=15 Physiol. Meas.\  
        \else\ifnum\jnl=16 Sov.\ Lightwave Commun.\ 
        \else\ifnum\jnl=17 J. Phys.\ D: Appl.\ Phys.\ 
        \else\ifnum\jnl=18 Supercond.\ Sci.\ Technol.\ 
        \else\ifnum\jnl=19 Semicond.\ Sci.\ Technol.\ 
        \else\ifnum\jnl=20 Nanotechnology\ 
        \else\ifnum\jnl=21 Meas.\ Sci.\ Technol.\  
        \else\ifnum\jnl=22 Plasma Sources Sci.\ Technol.\  
        \else\ifnum\jnl=23 Smart Mater.\ Struct.\  
        \else\ifnum\jnl=24 J.\ Micromech.\ Microeng.\ 
   \else Institute of Physics Publishing\  
   \fi\fi\fi\fi\fi\fi\fi\fi\fi\fi\fi\fi\fi\fi\fi 
   \fi\fi\fi\fi\fi\fi\fi\fi\fi} 
\let\abs=\beginabstract 

\let\endabs=\endabstract 
\def\today{\number\day\ \ifcase\month\or 
     January\or February\or March\or April\or May\or June\or 
     July\or August\or September\or October\or November\or 
     December\fi\space \number\year} 
\def\date{\ifppt\noindent\textfonts\rm  
     Date: \today\par\goodbreak\bigskip\fi} 
% 
% Physics Abstracts classification numbers 
% 
 
% 
 
% 
%%%%%%%%%%%%%%%%%%%%%%%%%%%%%%%%%%%%%%%%%%%%%%%%%%%%%%%%%%%% 
%                                                          % 
%  Sections, subsections, etc                              % 
%                                                          % 
%%%%%%%%%%%%%%%%%%%%%%%%%%%%%%%%%%%%%%%%%%%%%%%%%%%%%%%%%%%% 
% 
\def\section#1{\ifppt\ifnum\secno=0\eject\fi\fi 
    \subno=0\subsubno=0\global\advance\secno by 1 
    \gdef\labeltype{\seclabel}\ifnumbysec\countno=1\fi 
    \goodbreak\beforesecspace\nobreak 
    \noindent{\bf \the\secno. #1}\par\futurelet\next\sp@ce} 
\def\subsection#1{\subsubno=0\global\advance\subno by 1 
     \gdef\labeltype{\seclabel}% 
     \ifssf\else\goodbreak\beforesubspace\fi 
     \global\ssffalse\nobreak 
     \noindent{\it \the\secno.\the\subno. #1\par}% 
     \futurelet\next\subsp@ce} 
\def\subsubsection#1{\global\advance\subsubno by 1 
     \gdef\labeltype{\seclabel}% 
     \ifssf\else\goodbreak\beforesubsubspace\fi 
     \global\ssffalse\nobreak 
     \noindent{\it \the\secno.\the\subno.\the\subsubno. #1}\null.  
     \ignorespaces} 
% 
 
% 
%%%%%%%%%%%%%%%%%%%%%%%%%%%%%%%%%%%%%%%%%%%%%%%%%%%%%%%%%%%% 
%                                                          % 
%  Appendices                                              % 
%                                                          % 
%%%%%%%%%%%%%%%%%%%%%%%%%%%%%%%%%%%%%%%%%%%%%%%%%%%%%%%%%%%% 
% 
\def\numappendix#1{\ifappendix\ifnumbysec\countno=1\fi\else 
    \countno=1\figno=0\tabno=0\fi 
    \subno=0\global\advance\appno by 1 
    \secno=\appno\gdef\applett{A}\gdef\labeltype{\seclabel}% 
    \global\appendixtrue\global\numapptrue 
    \goodbreak\beforesecspace\nobreak 
    \noindent{\bf Appendix \the\appno. #1\par}% 
    \futurelet\next\sp@ce} 
\def\numsubappendix#1{\global\advance\subno by 1\subsubno=0 
    \gdef\labeltype{\seclabel}% 
    \ifssf\else\goodbreak\beforesubspace\fi 
    \global\ssffalse\nobreak 
    \noindent{\it A\the\appno.\the\subno. #1\par}% 
    \futurelet\next\subsp@ce} 
\def\@ppendix#1#2#3{\countno=1\subno=0\subsubno=0\secno=0\figno=0\tabno=0 
    \gdef\applett{#1}\gdef\labeltype{\seclabel}\global\appendixtrue 
    \goodbreak\beforesecspace\nobreak 
    \noindent{\bf Appendix#2#3\par}\futurelet\next\sp@ce} 
\def\Appendix#1{\@ppendix{A}{. }{#1}} 
\def\appendix#1#2{\@ppendix{#1}{ #1. }{#2}} 
\def\App#1{\@ppendix{A}{ }{#1}} 
\def\app{\@ppendix{A}{}{}} 
\def\subappendix#1#2{\global\advance\subno by 1\subsubno=0 
    \gdef\labeltype{\seclabel}% 
    \ifssf\else\goodbreak\beforesubspace\fi 
    \global\ssffalse\nobreak 
    \noindent{\it #1\the\subno. #2\par}% 
    \nobreak\subspace\noindent\ignorespaces} 
% 
%%%%%%%%%%%%%%%%%%%%%%%%%%%%%%%%%%%%%%%%%%%%%%%%%%%%%%%%%%%% 
%                                                          % 
%  Acknowledgments, notes added and foreign abstracts      % 
%                                                          % 
%%%%%%%%%%%%%%%%%%%%%%%%%%%%%%%%%%%%%%%%%%%%%%%%%%%%%%%%%%%% 
% 
\def\@ck#1{\ifletter\bigskip\noindent\ignorespaces\else 
    \goodbreak\beforesecspace\nobreak 
    \noindent{\bf Acknowledgment#1\par}% 
    \nobreak\secspace\noindent\ignorespaces\fi} 
\def\ack{\@ck{s}} 
\def\ackn{\@ck{}} 
\def\n@ip#1{\goodbreak\beforesecspace\nobreak 
    \noindent\smallfonts{\it #1}. \rm\ignorespaces} 
\def\naip{\n@ip{Note added in proof}} 
\def\na{\n@ip{Note added}} 
 
% 
%  \resume and \zus in Physics in Medicine and Biology only 
% 
 
% 
 
% 
%%%%%%%%%%%%%%%%%%%%%%%%%%%%%%%%%%%%%%%%%%%%%%%%%%%%%%%%%%%% 
%                                                          % 
%  Tables                                                  % 
%                                                          % 
%%%%%%%%%%%%%%%%%%%%%%%%%%%%%%%%%%%%%%%%%%%%%%%%%%%%%%%%%%% 
% 
 
% 
 
% 
 
\def\tablecont{\topinsert\global\advance\tabno by -1 
    \tablecaption{(continued)}} 
\def\tablecaption#1{\gdef\labeltype{\tablabel}\global\widefalse 
    \leftskip=\secindent\parindent=0pt 
    \global\advance\tabno by 1 
    \smallfonts{\bf Table \ifappendix\applett\fi\the\tabno.} \rm #1\par 
    \smallskip\futurelet\next\t@b} 
\def\t@b{\ifx\next*\let\next=\widet@b 
             \else\ifx\next[\let\next=\fullwidet@b 
                      \else\let\next=\narrowt@b\fi\fi 
             \next} 
\def\widet@b#1{\global\widetrue\global\notfulltrue 
    \t@bwidth=\hsize\advance\t@bwidth by -\secindent}  
\def\fullwidet@b[#1]{\global\widetrue\global\notfullfalse 
    \leftskip=0pt\t@bwidth=\hsize}                   
\def\narrowt@b{\global\notfulltrue} 
\def\align{\catcode`?=13\ifnotfull\moveright\secindent\fi 
    \vbox\bgroup\halign\ifwide to \t@bwidth\fi 
    \bgroup\strut\tabskip=1.2pc plus1pc minus.5pc} 
\def\endalign{\egroup\egroup\catcode`?=12} 
 
% 
% Use \L{#}, \R{#} and \C{#} to specify left, right or centred 
% columns immediately after \table. For example 
% \align\L{#}&&\L{#}\cr gives the preamble for a table with 
% all columns aligned left, \align\L{#}&\C{#}&\R{#}\cr 
% gives a table with 3 columns, the first aligned left, the second 
% centred and the third aligned right. 
% 

% 
%  Rules for tables \br at top and bottom 
%  \mr to separate headings from entries 
% 

% 
 
% 
% Definitions for centring headings over several columns 
% \centre{4}{Results for helium} will centre 
% Results for helium over four columns 
% \crule{4} will produce a rule centred over four columns 
% to go below a centred heading 
% 

% 
 
\catcode`?=13 
\def\lineup{\setbox0=\hbox{\smallfonts\rm 0}% 
    \digitwidth=\wd0% 
    \def?{\kern\digitwidth}% 
    \def\\{\hbox{$\phantom{-}$}}% 
    \def\-{\llap{$-$}}} 
\catcode`?=12 
% 
% Macros for two parts of a table of equal width side by side 
% \table{caption}[w] 
% \sidetable{first part}{second part} 
% \endtable 
% Use \table preamble for tables of 31picas width 
% 
\def\sidetable#1#2{\hbox{\ifppt\hsize=18pc\t@bwidth=18pc 
                          \else\hsize=15pc\t@bwidth=15pc\fi 
    \parindent=0pt\vtop{\null #1\par}% 
    \ifppt\hskip1.2pc\else\hskip1pc\fi 
    \vtop{\null #2\par}}}  
\def\lstable#1#2{\everypar{}\tempval=\hsize\hsize=\vsize 
    \vsize=\tempval\hoffset=-3pc 
    \global\tabno=#1\gdef\labeltype{\tablabel}% 
    \noindent\smallfonts{\bf Table \ifappendix\applett\fi 
    \the\tabno.} \rm #2\par 
    \smallskip\futurelet\next\t@b} 
\def\inctabno{\global\advance\tabno by 1} 
% 
%%%%%%%%%%%%%%%%%%%%%%%%%%%%%%%%%%%%%%%%%%%%%%%%%%%%%%%%%%%% 
%                                                          % 
%  Figures                                                 % 
%                                                          % 
%%%%%%%%%%%%%%%%%%%%%%%%%%%%%%%%%%%%%%%%%%%%%%%%%%%%%%%%%%%% 
% 
 
% 
 
% 
\def\figure#1{\figc@ption{#1}\bigskip} 
\def\figc@ption#1{\global\advance\figno by 1\gdef\labeltype{\figlabel}% 
   {\parindent=\secindent\smallfonts\hang 
    {\bf Figure \ifappendix\applett\fi\the\figno.} \rm #1\par}} 
% 
%%%%%%%%%%%%%%%%%%%%%%%%%%%%%%%%%%%%%%%%%%%%%%%%%%%%%%%%%%%% 
%                                                          % 
%  Reference lists                                         % 
%                                                          % 
%%%%%%%%%%%%%%%%%%%%%%%%%%%%%%%%%%%%%%%%%%%%%%%%%%%%%%%%%%%% 
% 
\def\refHEAD{\goodbreak\beforesecspace 
     \noindent\textfonts{\bf References}\par 
     \let\ref=\rf 
     \nobreak\smallfonts\rm} 
\def\references{\refHEAD\parindent=0pt 
     \everypar{\hangindent=18pt\hangafter=1 
     \frenchspacing\rm}% 
     \secspace} 
\def\rf#1{\par\noindent\hbox to 21pt{\hss #1\quad}\ignorespaces} 
% 
 
% 
 
% 
% reference to a journal article in numerical system 
% 
 
% 
% reference to a book or report in numerical system 
% 
 
% 
%%%%%%%%%%%%%%%%%%%%%%%%%%%%%%%%%%%%%%%%%%%%%%%%%%%%%%%%%%%% 
%                                                          % 
%  Theorems, lemmas, etc                                   % 
%                                                          % 
%%%%%%%%%%%%%%%%%%%%%%%%%%%%%%%%%%%%%%%%%%%%%%%%%%%%%%%%%%%% 
% 

% 
% NB \note#1 is used to give a Note (as opposed to a paper or letter) 
% in PMB therefore use commands \notes#1 for numbered Note 
% instead of \note  
% 

% 
\catcode`\@=12 
% 
% Parameter values for `Preprint' style  
% 
 
% 
% Parameter values for `Journal' style  
% 
\def\jnlstyle{\pptfalse\headsize{14}{18}% 
\textsize{10}{12}% 
\smallsize{8}{10} 
\textind=16pt} 
% 
% Parameter values for `Eleven point' style  
% 
 
% 
% Parameter values for `Large size' style  
% 
 
% 
\parindent=\textind 

\input xref
%\pptstyle
\firstpage=1
\pageno=1

\jnlstyle

\jl{1}

\overfullrule=0pt
\letter{Lattice two-point functions and conformal invariance}[Letter to 
the Editor]

\author{Malte Henkel and Dragi Karevski}[Letter to the Editor]

\address{Laboratoire de Physique des Mat\'eriaux\footnote{\dag}{
Unit\'e Mixte de Recherche  CNRS No~7556},  Universit\'e Henri 
Poincar\'e (Nancy~I), 
BP~239, F--54506  Vand\oe uvre l\`es Nancy Cedex, France}

\abs
A new realization of the conformal algebra is studied which mimics the
behaviour of a statistical system on a discrete albeit infinite lattice. 
The two-point function is found from the requirement that it transforms 
covariantly under this realization. The result is in agreement with explicit 
lattice calculations of the $(1+1)D$ Ising model and the $d-$dimensional 
spherical model. A hard core is found which is not present in the continuum.
For a semi-infinite lattice, profiles are also obtained.
\endabs
\vglue 1.5cm
\date

Since the pioneering work of Polyakov (1970), 
conformal invariance has become
a powerful tool in the description and understanding of critical phenomena,
in particular for two-dimensional systems (Belavin {\it et al\/} 1984). 
For example critical exponents and the $n-$point correlation 
functions at criticality can be exactly determined and
a classification of the universality classes is obtained, see e.g.
Christe and Henkel (1993) and di Francesco {\it et al\/} (1997) for an
introduction.

The basis of this theory is the assertion that there is a certain class of
scaling operators, called (quasi)primary, which transform covariantly under
(global) conformal transformations. For notational simplicity, 
we shall work in a
two-dimensional setting throughout, but the generalization to higher
spatial dimensions will be immediate. For the same reason, we restrict
attention to scalar scaling operators. The generators of the conformal algebra
may be written down in the form ${\ell}_n = - \hat{X}^{n+1} \hat{P}$, where
$\hat{X}$ and $\hat{P}$ are operators which satisfy the commutation relation
$[\hat{P},\hat{X}]=1$ (we neglect here the terms involving the central
charge). Usually, an underlying continuum theory is assumed and then the
choice of realization 
$$
\hat{X} = z \;\; , \;\; \hat{P} = {\partial \over \partial z} 
\eqno(1)
$$
is natural. However, other realizations are perfectly possible. For example,
in the context of $2D$ turbulence, new realizations of the conformal algebra
which yield logarithmic two-point functions, are needed 
(Rahimi Tabar {\it et al\/} 1997, Flohr 1996). Here, we shall consider yet
another realization which mimics the behaviour of a system defined on a
discrete lattice. Field theories on a discrete (space-time) lattice are 
presently under active study, 
e.g. Kauffman and Noyes (1996), Winitzki (1997) and
Faddeev and Volkov (1997) and references therein. 
Following an idea exploited in connection with
Schr\"odinger invariance (Henkel and Sch\"utz 1994), we shall work with
$$
\hat{X} = {1 \over \cosh {a\over 2}\partial_z} \, z \;\; , \;\;
\hat{P} = {2\over a} \sinh \left( {a\over 2} \partial_z \right)
\eqno(2)
$$
where $a$ is a free parameter which we shall interpret as a lattice constant. 
Comparing the realizations (1) and (2), we see that the usual infinitesimal
translation operator ${\ell}_{-1}^{(a=0)} f(z)= -\partial_z f(z)$ is replaced 
by the symmetric discrete difference
${\ell}_{-1}^{(a)}f(z)= -\partial_z^a f(z) := -\frac{1}{a}
\left[f\left(z+\frac{a}{2}\right)-f\left(z-\frac{a}{2}\right)\right]$.

For a better geometric understanding of these transformations, we briefly
consider the finite transformations generated by the ${\ell}_n$. These are 
given by the functions $S(\epsilon,z) = \exp(\epsilon \ell_n)z$, 
together with the
initial condition $S(0,z)=z$. They may be obtained by solving the equation
$$
{\partial S \over \partial \epsilon}(\epsilon,z) = \ell_n S(\epsilon,z)
\eqno(3)
$$
In the case of the continuum realization (1), one has 
$S(\epsilon,z)=z-\epsilon$ for
translations ($n=-1$) and $S(\epsilon,z)=z e^{-\epsilon}$ 
for dilatations ($n=0$),
respectively. However, it can be easily checked that 
the same solutions of~(3) also
apply for the `lattice' realization (2). The lattice is mapped onto itself,
provided $\epsilon$ is restricted to discrete values, e.g. $\epsilon=m a$ 
($m$ integer) for translations and $\epsilon=-\ln p$ ($p$ a positive integer) 
for dilatations. Discrete dilatations also 
appear naturally in connection with the generators of (multi)fractal sets
(Erzan and Eckmann 1997) and the critical behaviour of certain aperiodic
quantum chains (Karevski and Turban 1996), see also Sornette (1997). 

We now focus on the calculation of the two-point correlation function
$F(r_{1},t_{1};r_{2},t_{2})=
\langle\phi_1(r_{1},t_{1})\phi_2(r_{2},t_{2})\rangle$. We assume that the
`space' direction is discretized and the `time' direction is continuous,
that the system is infinite in both directions and 
invariant under the global conformal group generated by $\ell_{\pm1,0}$
and $\bar{\ell}_{\pm1,0}$. Here the $\phi_i$ are (scalar) 
scaling operators of scaling
dimensions $x_i$. Following the procedure which is standard in the 
continuum (Polyakov 1970), we require $F$ to transform covariantly
under the `lattice' realization (2). 

{}From translation invariance in both space and time directions, it appears
physically obvious to take 
$F(r_{1},t_{1};r_{2},t_{2})=F(r,t)$ with $r=r_{2}-r_{1}$ and $t=t_{2}-t_{1}$,
since this agrees with the continuum limit $a\to 0$ (more precisely, the
equation $(\partial_{r_1}^{a} + \partial_{r_2}^{a})F=0$ has two solutions,
only one of which has the expected $a\to 0$ limit (Henkel and Sch\"utz 1994)).
{}From rotational invariance\footnote{\ddag}{For brevity, this infinitesimal 
transformation is called rotation by analogy with the real infinitesimal 
rotation to which it reduces for $a\rightarrow 0$.},
$F$ should satisfy the differential equation
$i(\ell_0 - \bar{\ell}_0)F(r,t) = \left[-t\partial_{r}^{a}+
(\cosh({a\over 2}\partial_r))^{-1} r\partial_{t}\right]F(r,t)=0$.
Fourier transformation gives
$$
\left[\frac{2}{a}\sin\left(\frac{a}{2}k\right)\partial_{\omega}
-\frac{\omega}{\cos\left(\frac{a}{2}k\right)}\partial_{k}\right]
\widetilde{F}(k,\omega)=0
\eqno(4)
$$
where $\widetilde{F}(k,\omega)$ is the Fourier transform of $F(r,t)$. 
The general solution of (4) is 
$\widetilde{F}(k,\omega)=\widetilde{\cal F}
\left(\omega^{2}+\left[\frac{2}{a}\sin(ak/2)\right]^{2}\right)$, where
$\cal F$ is an arbitrary function 
(in the continuum limit, $a\to 0$, one recovers the usual form 
$\widetilde{\cal F}(\omega^{2}+k^{2})$). 

Here a comment is in order. The argument of the function $\cal F$ is
related to the Casimir operator $C$ of the subalgebra generated by 
translations and rotations. Dilatation invariance fixes $C=0$, or 
equivalently, yields the dispersion relations $E^2 = k^2$ for the continuum
and $E^2 = 4/a^2 \sin^2(ak/2)$ for the `lattice' realization, respectively. 
Using the `lattice' realization (2) automatically leads to the lattice 
dispersion relation.
We can therefore expect the hypothesis of covariance under the 
`lattice' realization (2) to apply to systems
which satisfy that dispersion relation on a discrete lattice with lattice
constant $a$.  
Model examples for this situations are provided by systems with an underlying 
field theory satisfying this dispersion relation, see below.

Finally, covariance under dilatations, 
generated by $\ell_0 + \bar{\ell}_0$,
gives in Fourier space
$$
\left[\partial_{\omega}\omega+\frac{2}{a}
\frac{1}{\cos\left(\frac{ak}{2}\right)}
\partial_{k}\sin\left(\frac{ak}{2}\right)\right]\widetilde{F}(k,\omega)=
\left(x_{1}+x_{2}\right)\widetilde{F}(k,\omega)\; ,
\eqno(5)
$$
(where the differential operators act on everything to the right)
and the solution is 
$\widetilde{F}(k,\omega)=\widetilde{F_{0}}\left[\omega^{2}+
\left(\frac{2}{a}\sin(ak/2)\right)^{2}\right]^{(x_1+x_2-2)/2}$,  
where $\widetilde{F_{0}}$ is a constant. 
Finally, a straightforward calculation
shows (see Henkel and Sch\"utz 1994) that covariance under 
$\ell_1, \bar{\ell}_1$
merely adds the extra condition $x_1 = x_2 =: x_{\phi}$.
The two-point function is
$$
F(r,t)=A \, \delta_{x_1,x_2}\, 
a^{-2x_{\phi}}\int_{0}^{\infty}\!{\rm d}v\;v^{-x_{\phi}-\frac{1}{2}}
\;\exp\left(-\frac{\tau^{2}}{v}-v\right)
\;{\rm I}_{\rho}(v)
\eqno(6)
$$
with $\tau^{2}=t^{2}/(2a^{2})$, $\rho=r/a$ is a positive integer, 
${\rm I}_{\rho}(v)$ is a modified Bessel
function and $A$ is a normalization constant. 
In fact, this can be written explicitly 
in terms of hypergeometric functions ${}_{1}F_{2}$ 
(Prudnikov {\it et al\/} 1992) and reads 
$F(r,t)=A\delta_{x_1,x_2}a^{-2x_{\phi}}\Psi$, where
$$\eqalign{
\!\!\!\!\!\!\!\!\!\!\!\!\!\!\!\!\!\!\!\!\!\!\!\!\!\!\!\!
\Psi=\frac{2^{x_{\phi}-
\frac{1}{2}}}{\sqrt{\pi}}
\frac{\Gamma(\rho+\frac{1}{2}-x_{\phi})\Gamma(x_{\phi})}{\Gamma(\rho+
\frac{1}{2}+x_{\phi})}
~_{1}F_{2}(x_{\phi}\;;\;x_{\phi}+\frac{1}{2}-\rho\;,\;x_{\phi}+
\frac{1}{2}+\rho\;;\;2\tau^{2})\cr
\!\!\!\!\!\!\!\!\!\!\!\!\!\!\!\!\!\!\!
+2^{-\rho}\tau^{2(\rho+\frac{1}{2}-x_{\phi})}
\frac{\Gamma(x_{\phi}-\frac{1}{2}-
\rho)}{\Gamma(1+\rho)}
~_{1}F_{2}(\rho+\frac{1}{2}\;;\;\frac{3}{2}-x_{\phi}+\rho\;,\;
1+2\rho\;;\;2\tau^{2})
\;.\cr}\eqno(7)
$$
A particularly simple form is found for the correlation purely in the `space' 
direction, along the discrete lattice, that is $t=0$
$$
F(r,0)=F_0 \, \delta_{x_1, x_2} \, a^{-2x_{\phi}} \, 
\frac{\Gamma(\rho+{1\over2}-x_{\phi})}{\Gamma(\rho+{1\over2}+x_{\phi})}\; .
\eqno(8)
$$
In the limit $\rho = r/a\gg 1$ this reduces to the standard continuum form 
$F(r)=F_0 r^{-2x_{\phi}}$, as it should be. We also remark that the power-law
dependence on $a$ is consistent with the interpretation of $a$ as a length. 

A new feature of the `lattice' realization is a {\it hard core} in the 
two-point function. This effect is absent in the continuum limit $a\to 0$.
To see this, note that for $\frac{1}{2}+ n \leq x_{\phi} < \frac{1}{2}+n+1$
with $n=0,1,2,\ldots$, the position $\rho_s$ of the first singularity of
$F(r,0)$ is in the interval $n\leq \rho_s < n+1$ 
(for $x_{\phi} < \frac{1}{2}$,
there is no singularity of $F(r,0)$ for positive $r$). For $\rho \leq \rho_s$,
$F(r,0)$ is no longer defined. The system behaves as
if the scaling operator $\phi$ would contain a hard sphere of diameter
${\cal D} =(x_{\phi}-\frac{1}{2})a$, thus preventing an approach closer 
than $\rho=\rho_s\geq n$. For $x_{\phi} \leq \frac{1}{2}$, the operator $\phi$
does not contain such a hard sphere at all. For larger values of $x_{\phi}$,
a hard sphere is always present, but for $x_{\phi}< \frac{3}{2}$, its
diameter $\cal D$ is still smaller than the lattice constant $a$. 
On the other hand, the scaling operator corresponding to the order parameter 
$\sigma$ should be as local as possible. 
If we would take as locality criterion
the absence of a hard sphere at all, this suggests that its scaling dimension
$x_{\sigma}\leq\frac{1}{2}$, but if it enough that two order parameter
operators can be defined on neighbouring sites with distance $a$, 
eq.~(8) suggests that $x_{\sigma} \leq \frac{3}{2}$.

As already noted above, eq.~(6) should be correct for models with 
an underlying free field theory
satisfying the dispersion relation $E=(2/a)|\sin(ak/2)|$. Examples of these
are the spherical model and the even sector of the $(1+1)D$ Ising model.
Consider first a $d$-dimensional spherical model,  
continuous in the $d-1$ directions labeled by a vector
position ${\bf r}_{\perp}$ and discrete in 
the direction $r_{\parallel}$, with
$a=1$. The spin-spin correlation function at the critical temperature 
$T_{c}$ is (Berlin and Kac 1952) 
$$
\!\!\!\!\!\!\!\!\!\!\!\!\!\!\!\!\!\!
C(r_{\parallel},{\bf r}_{\perp})=
\frac{k_{B}T_{c}}{(2\pi)^{d}}\int_{-\pi}^{\pi}\dots\int_{-\pi}^{\pi}
{\rm d}\phi_{1}\dots{\rm d}\phi_{d}\; 
\frac{\cos({\bf r}_{\perp}{\Phi}_{\perp}+r_{\parallel}\phi_{\parallel})}
{2Jd-2J(\cos\phi_{1}+\dots+\cos\phi_{d})}\qquad
\eqno(9)
$$
where $J$ is the coupling constant and ${\Phi}_{\perp}$ is the vector
$(\phi_1,\ldots,\phi_{d-1})$. Using the expansion 
$\cos\phi_{i}\simeq 1-\phi_{i}^{2}/2$ in the $d-1$ continuous directions, 
we get in a standard fashion, for $2<d<4$
$$
\!\!\!\!\!\!\!\!
C(r_{\parallel},{\bf r}_{\perp})=\frac{k_{B}T_{c}}{2J(2\pi)^{(d-1)/2}}
\int_{0}^{\infty}\!{\rm d}u\;u^{-{{d-2}\over2}-\frac{1}{2}}
\;\exp\left(-\frac{r_{\perp}^{2}}{2u}-u\right)
\;{\rm I}_{r_{\parallel}}(u)\; ,
\eqno(10)
$$
where $r_{\perp}^{2}=\sum_{i=1}^{d-1}r_{i}^{2}$ and $r_{\parallel}$ are both
integers. With the identification
$x_{\phi}=x_{\sigma}=\beta/\nu=(d-2)/2$ for the spherical model, 
we get exactly the same form as eq.~(6).

We turn now to the $(1+1)D$ Ising model. In fact, a test of
eq.~(8) is easier to perform in the strongly anisotropic limit, which implies
a continuum limit in the `time' direction while keeping the `space'
direction discrete, maps the
$2D$ Ising model onto the Ising quantum chain in a transverse field. 
Then the two-point correlation
function for the energy density operator $\varepsilon$ is (Pfeuty 1970)
$$
C_{\varepsilon}(\rho,0)= \frac{1}{4\rho^{2}-1}
\eqno(11)
$$
while from eq.~(8), with $x_{\varepsilon}=1$ for the 
scaling dimension of the energy density,
we expect up to normalization $F_{\varepsilon}(\rho,0)=1/(4\rho^{2}-1)$.
On the other hand, the spin-spin correlation function is not of the form~(8), 
but for $\rho\gg 1$ one has (Pfeuty 1970)
$$
{C_{\sigma}(\rho,0)}\simeq {C_{\sigma,0}} \,  
\rho^{-1/4}\left[1-\frac{1}{64}\rho^{-2}\right]\; ,
\eqno(12)
$$
while our formula~(8), with $x_{\sigma}=1/8$, leads to
${F_{\sigma}(\rho,0)}\sim 
\rho^{-1/4}\left[1-1/(48\rho^{2})\right]$.
This discrepancy between odd and even sectors of the Ising model
is not surprising. 
Indeed, {\it only} the even sector alone is described by a free fermionic 
field. However, the order parameter intertwines between the
even and odd symmetry sectors of the model and does {\it not\/} correspond to
a free field with a lattice dispersion relation $E=(2/a)|\sin(ak/2)|$, in
contrast to the requirements in the derivation of (8). 

Finally, we consider briefly the case of a semi-infinite geometry, 
with the boundary perpendicular to the discretization axis, 
so that we have for the position coordinates 
$-\infty<{\bf r}_{\perp}<\infty$ and $0<r_{\parallel}<\infty$.
We consider in this case a
non-vanishing profile
of a scaling operator $\phi$, $C({\bf r}_{\perp},r_{\parallel})=
\langle \phi({\bf r}_{\perp},r_{\parallel})\rangle$.
{}From translation invariance along the perpendicular directions, 
$C({\bf r}_{\perp},r_{\parallel})=F(r_{\parallel})$. 
Using a Laplace transformation
$$
F(r_{\parallel})=\int_{0}^{\infty}\e^{-r_{\parallel}s}f(s) {\rm d}s\; ,
\eqno(13)
$$
covariance of $F(r_{\|})$ under dilatations (generated by $\ell_0$) leads to,
for $x_{\phi}>0$
$$
\frac{2}{a}\frac{1}{\cosh\left(\frac{as}{2}\right)}\partial_{s}\left(
\sinh\left(\frac{as}{2}\right)f(s)\right)=x_{\phi}f(s)\; ,
\eqno(14)
$$
with the solution
$f(s)=A\left[2 a^{-1}\sinh\left(\frac{as}{2}\right)\right]^{x_{\phi}-1}$,
where $A$ is a constant. We find
$$
F(r_{\parallel})=F_0 \, a^{-x_{\phi}} \, 
\frac{\Gamma\left({r_{\parallel}\over a}+
{1\over2}-\frac{x_{\phi}}{2}\right)}
{\Gamma\left({r_{\parallel}\over a}+
{1\over2}+\frac{x_{\phi}}{2}\right)}\; 
\eqno(15)
$$
which is exactly the same result as in eq.~(8), but for the replacement of
$x_{\phi}$ by $x_{\phi}/2$.  In the continuum limit,
$r_{\parallel}/a\gg 1$, we recover the Fisher-de Gennes~(1978)  result
$F(r_{\parallel})\sim r_{\parallel}^{-x_{\phi}}$. We observe an analogous
hard core effect as found for the two-point function for the infinite system. 

Summarizing, we have studied a new realization of the conformal algebra which 
mimics a system defined on a discrete lattice. From the covariance under global
conformal transformations, the two-point function was obtained and found to be
in agreement with explicit results in the spherical model and the even sector 
of the $(1+1)D$ Ising model. 
It displays a hard core not present in the continuum limit.

\references

AA Belavin, AM Polyakov and AB Zamolodchikov 1984,
Nucl. Phys. {\bf B241}, 333

TH Berlin and M Kac 1952, Phys. Rev. {\bf 86}, 821
 
P Christe and M Henkel 1993, {\it Introduction to
Conformal Invariance and its Applications to Critical Phenomena}, 
Springer (Heidelberg)

A Erzan and JP Eckmann 1997, Phys. Rev. Lett. {\bf 78}, 3245

P di Francesco, P Mathieu and D S\'en\'echal 1997, 
{\it Conformal Field Theory}, Springer (Heidelberg)

LD Faddeev and AYu Volkov, {\it Algebraic Quantization of Integrable Models
in Discrete Space-time}, preprint hep-th/9710039

ME Fisher and PG de Gennes 1978, C. R. Acad. Sci. Paris {\bf B287}, 207

MAI Flohr 1996, Nucl. Phys. {\bf B482}, 567

M Henkel and GM Sch\"utz 1994, Int. J. Mod. Phys. 
{\bf B8}, 3487

D Karevski and L Turban 1996, J. Phys. {\bf A29}, 3461

LH Kauffman and HP Noyes 1996, Phys. Lett. {\bf 218A}, 139

P Pfeuty 1970, Ann. of Phys. {\bf 57}, 79

AM Polyakov 1970, Sov. Phys. JETP Lett. {\bf 12}, 381

AP Prudnikov, YuA Brychkov and OI Marichev 1992, {\it Integrals and Series}, 
Vol. 2, eq. (2.15.6.6), Gordon and Breach (New York)
 
MR Rahimi Tabar, A Aghamohammadi and M Khorrami 1997, Nucl. Phys. {\bf B497},
555

D Sornette 1997, {\it Discrete scale invariance and complex dimensions}, 
preprint cond-mat/9707012, submitted to Phys. Rep. 

S Winitzki 1997, Phys. Rev. {\bf D55}, 6374 

\end{document}